\documentstyle[epsfig]{aa}

\begin{document}

%\thesaurus{}
\title{The very early afterglow powered by ultra-relativistic mildly magnetized outflows}
\author{Y. Z. Fan\inst{1,2}, D. M. Wei\inst{1,2} and C. F. Wang\inst{1,2}}

\offprints {Y. Z.~Fan}

\mail{yzfan@pmo.ac.cn}

\institute{Purple Mountain Observatory, Chinese Academy of
Sciences, Nanjing, 210008, China \and National Astronomical
Observatories, Chinese Academy of Sciences, Beijing, 100012,
China}
\date{Received date 19 April 2004/ Accepted date 18 May 2004}
\titlerunning{Reverse shock emission powered by magnetized ejecta}

\abstract{In the Poynting Flux-dominated outflow (the initial
ratio of the electromagnetic energy flux to the particle energy
flux $\sigma_0\gg1$) model for Gamma-ray bursts, particularly the
$\gamma-$ray emission phase, nearly half of the internally
dissipated magnetic energy is converted into the $\gamma-$ray
energy emission and the rest is converted into the kinetic energy
of the outflow. Consequently, at the end of the $\gamma-$ray
burst, $\sigma$ decreases significantly ($\sigma\sim 1$ or even
smaller). We numerically investigate the very early reverse shock
emission powered by such mildly magnetized outflows interacting
with medium---uniform interstellar medium (ISM) or stellar wind
(WIND). We show that for $\sigma\sim0.05-1$ and typical parameters
of Gamma-ray bursts, both the ISM-ejecta interaction and the
WIND-ejecta interaction can power very strong optical emission
($m_{\rm R}\sim 10-12{\rm th}$ magnitude or even brighter).
Similar to the very early afterglow powered by the non-magnetized
ejecta interacting with the external medium, the main difference
between the ISM-ejecta interaction case and the WIND-ejecta
interaction case is that, before the reverse shock crosses the
ejecta, the R-band emission flux increases rapidly for the former,
but for the latter it increases only slightly.

At the very early stage, the ejecta are ultra-relativistic. Due to
the beaming effect, the random magnetic field generated in shocks
contained in the viewing area is axisymmetric, unless the line of
sight is very near the edge of ejecta. The formula $\Pi_{\rm
net}\approx 0.60b^2/(1+b^2)$ (where $b$ is the ratio of the
ordered magnetic field strength to that of random one) has been
proposed to describe the net linear polarization of the
synchrotron radiation coming from the viewing area. For
$\sigma\sim 0.05-1$, the ordered magnetic field dominates over the
random one generated in the reverse shock (As usual, we assume
that a fraction $\epsilon_{\rm B}\sim 0.01$ of the thermal energy
of the reverse shock has been converted into the magnetic energy),
the high linear polarization is expected. We suggest that the
linear polarization detection of the early multi-wavelength
afterglow is required to see whether the outflows powering GRBs
are magnetized or not.

}

\maketitle

\section{Introduction}
It is widely accepted that $\gamma$-ray bursts (GRBs) are powered
by the dissipation of energy in a highly relativistic wind, driven
by gravitational collapse of a massive star into a neutron star or
a black hole (see M\'{e}sz\'{a}ros 2002 for a recent review). As
the observed emission is powered at a distance far from the
central source, key questions remain unanswered. One of them is
the Gamma-ray burst engines (see Cheng \& Lu 2001 for a review).
In the standard fireball model, the Gamma-ray burst are powered by
the collisions of baryon dominated shells with variable Lorentz
factors (Paczy\'{n}ski \& Xu 1994; Rees \& M\'{e}sz\'{a}ros 1994).
However, Poynting flux-driven outflows from magnetized rotators is
another plausible explantation and there have been various
implementations of this concept (Usov 1992, 1994; Thompson 1994;
Blackman, Yi \& Field 1996; Katz 1997; M\'{e}sz\'{a}ros \& Rees
1997). The Poynting flux model is in light of the following facts
(see Zhang \& M\'{e}sz\'{a}ros (2004) for a recent review): The
Poynting flux outflow can transport a large amount of energy
without carrying many baryons. It can alleviate the inefficiency
problem of the internal shock model and can also alleviate the
magnetic field amplification problem in GRBs and afterglows. It
provides the possibility of achieving narrow ``peak energy''
distributions. In particular, the Poynting flux model provides the
most natural explanation so far for the very high linear
polarization during the $\gamma$-ray emission phase of GRB 021206
(e.g. Coburn \& Boggs 2003; Lyutikov, Pariev \& Blandford 2003;
Granot 2003; However, see Rutledge \& Fox (2004) for argument),
although other alternative explanations remain (e.g. Shaviv \& Dar
1995; Waxman 2003).

In the past several years, many publications have focused on the
dissipation (the energetic non-thermal $\gamma-$ray emission) as
well as the acceleration of the Poynting flux outflow (e.g. Usov
1994; Thompson 1994; Smolsky \& Usov 1996; M\'{e}sz\'{a}ros \&
Rees 1997; Lyutikov \& Blackman 2001; Spruit, Daigne \& Drenkhahn
2001; Drenkhahn 2002; Drenkhahn \& Spruit 2002). In this paper we
turn to investigate the very early reverse shock emission powered
by such a magnetized outflow, just as Sari \& Piran (1999) and
M\'{e}sz\'{a}ros \& Rees (1999) have done for the baryon dominated
fireball. One thing inciting us to do this is that modeling the
very early afterglow of GRB 990123 and GRB 021211 suggests the
reverse shock emission region is magnetized (Fan et al. 2002;
Zhang, Kobayashi \& M\'{e}sz\'{a}ros 2003).

In the ``internal'' magnetic dissipation model, GRBs are powered
by the magnetic energy dissipation at a radius $r\sim
10^{13}-10^{14}{\rm cm}$. At the end of the $\gamma-$ray burst, a
significant fraction of magnetic energy has been dissipated by
magnetic reconnection or other processes. Nearly half of the
dissipated magnetic energy has been converted into the
$\gamma-$ray emission and the rest has been converted into the
kinetic energy of the outflow (see Spruit $\&$ Drenkhahn 2003 for
a recent review). Consequently, $\sigma$ decreases significantly
($\sim 1$ or even smaller). At a much larger radius, where the
outflow begins to be decelerated significantly, the reverse shock
emission (the very early afterglow) is expected; this is what we
focus on.

At the final stage of the preparation of this manuscript, a paper
by Zhang \& Kobayashi (2004) appeared. In that paper, the very
early reverse shock emission from an arbitrary magnetized ejecta
has been analytically investigated.

\section{The mild-magnetized outflow}
As mentioned before, there are lots of publications focused on the
acceleration of the magnetized outflow. One of them is Drenkhahn
(2002), in which part of the magnetic energy coupled with the
outflow is dissipated internally by reconnection and the Lorentz
factor of the flow increases steadily with radius ($\Gamma\propto
r^{1/3}$). Here we do not discuss that topic further and just take
the numerical example presented in Drenkhahn $\&$ Spruit (2002) as
the starting point of our calculation: at the end of the prompt
$\gamma-$ray emission phase, the bulk Lorentz factor of the
outflow is $\eta \sim 300$; the ratio of the electromagnetic
energy flux to the particle energy flux, $\sigma\sim 0.05-1$ (In
principle, much lower $\sigma$ is possible, for which the reverse
shock emission is similar to that of the usual fireball, which is
beyond our interest); the total kinetic energy (including the
magnetic energy) is of the order of the typical $\gamma-$ray
emission energy, i.e., $E_{\rm kin}\sim 10^{53}{\rm ergs}$.

\section{The reverse shock emission}
\subsection{The dynamical evolution of ejecta}
Generally, the dynamical evolution of the ejecta can be divided
into two phases---(i) Before the reverse shock crosses the ejecta,
i.e., $R<R_{\rm cro}$ ($R$ is the radial coordinate in the burster
frame; $R_{\rm cro}$ is the radius at which the reverse shock
crosses the ejecta); at that time two shocks exist. The dynamical
evolution of the ejecta is governed by the jump condition of
shocks (e.g., Blandford \& McKee 1976; Sari \& Piran 1995); (ii)
After the reverse shock crosses the ejecta, i.e., $R>R_{\rm cro}$,
in this case only the forward shock exists. The hydrodynamical
evolution can be calculated by taking the generic dynamical model
of GRB remnants (e.g., Huang, Dai \& Lu 1999; Huang et al. 2000;
Feng et al. 2002).

\subsubsection{The dynamical evolution for $R<R_{\rm cro}$}

Similar to Sari \& Piran (1995), the dynamical evolution of the
ejecta is obtained by solving the jump condition for strong
shocks. For the ejecta interacting with the external medium, there
are two shocks formed, one is the forward shock expanding into the
medium, the other is the reverse shock penetrating the ejecta.
There are four regions in this system: (1) The un-shocked medium;
(2) The shocked medium; (3) The shocked ejecta material; (4) The
un-shocked ejecta material. The medium is at rest relative to the
observer. The bulk Lorentz factors $\Gamma_{\rm j}$ ($j=1,4$,
$\Gamma_4\equiv \eta$) and the corresponding velocities
$\beta_{\Gamma_{\rm j}}=(1-1/\Gamma_{\rm j}^2)^{1/2}$ are measured
by the observer. Thermodynamic quantities: $n_{\rm j}$, $p_{\rm
j}$, $e_{\rm j}$, $B'_{\rm j}$ (particle number density, pressure,
internal energy density, magnetic field strength) are measured in
the fluids' rest frame (we assume the un-shocked ejecta and medium
are cold, i.e., $e_4=e_1=0$), so is the $p_{\rm B,i}$ (the
magnetic pressure). The equations governing the forward shock are
(Blandford $\&$ Mackee 1976)
\begin {equation}
n_2/n_1=4\Gamma_2+3, e_2/n_2=(\Gamma_2-1)m_p c^2.
\end {equation}

Below, following Kennel \& Coroniti (1984, hereafter KC84) we try
to derive the ($90^{\rm o}$) shock jump condition governing the
reverse shock with the MHD conservation laws  by assuming the
magnetic field frozen in the outflow is nearly toroidal (KC84):
\begin{equation}
n_4u_4=n_3u_3,
\end{equation}
\begin{equation}
\gamma_4 \mu_4+{EB_4\over 4\pi n_4u_4}=\gamma_3 \mu_3+{EB_3\over
4\pi n_4u_4},
\end{equation}
\begin{equation}
\mu_4u_4+{P_4\over n_4 u_4}+{B_4^2\over 8\pi
n_4u_4}=\mu_3u_3+{P_3\over n_4 u_4}+{B_3^2\over 8\pi n_4u_4},
\end{equation}
where $B_{\rm i}$ and $E$ denote the shock frame magnetic and
electric fields respectively, $E=u_4B_4/\gamma_4=u_3B_3/\gamma_3$.
$\gamma_{\rm i}$ ($i=3,~4$) is the Lorentz factor of the fluid
measured in the reverse shock frame, $u_{\rm i}^2=\gamma_{\rm
i}^2-1$, $\beta_{\rm i}=u_{\rm i}/\gamma_{\rm i}$. $\mu$ is the
specific enthalpy, which is defined by $\mu_{\rm i}=m_{\rm
p}c^2+\hat{\gamma}_{\rm i}P_{\rm i}/[(\hat{\gamma}_{\rm
i}-1)n_{\rm i}]$ for a gas with an adiabatic index
$\hat{\gamma_{\rm i}}$.

Solving equation (3) for $\mu_3$ and inserting the resulting
expression into equation (4) leads to
\begin{eqnarray}
\sigma (Y^2-1)&+&{2[1+\sigma(1-Y)]\over u_3\gamma_3}{u_4\over
\gamma_4}[u_3^2+{\hat{\gamma}_3-1\over
\hat{\gamma}_3}]-2({u_4\over
\gamma_4})^2\nonumber\\
&-&{2P_4\over n_4\mu_4\gamma_4^2}-{u_4\over
u_3}{2(\hat{\gamma}_3-1)m_{\rm p}c^2\over
\hat{\gamma}_3\mu_4\gamma_4^2}=0,
\end{eqnarray}
where $\sigma\equiv {B_4^2/[4\pi n_4\mu_4 \gamma_4^2]}$ and
$Y\equiv {\gamma_3 u_4\over \gamma_4 u_3}$. With equation (3), the
downstream pressure $P_3$ can be calculated as follows:
\begin{equation}
P_3={\hat{\gamma}_3-1\over \hat{\gamma}_3}[{\gamma_4 \over
\gamma_3}(1+\sigma(1-Y))-{m_{\rm p}c^2\over \mu_4}]n_3\mu_4.
\end{equation}

For $\gamma_4 \gg 1$ and $P_4=0$, equation (5) can be arranged
into (Assuming $2(\hat{\gamma}_3-1)/u_3\hat{\gamma}_3\gamma_4$ can
be ignored)
\begin{eqnarray}
&&{(1+\sigma)(2-\hat{\gamma}_3)\over
\hat{\gamma}_3}u_3^4+[{\hat{\gamma}_3-4\over
4\hat{\gamma}_3}\sigma^2-{\hat{\gamma}_3^2-2\hat{\gamma}_3+2\over
\hat{\gamma}_3^2}\sigma\nonumber\\
&&~~~~~-({\hat{\gamma}_3-1\over
\hat{\gamma}_3})^2]u_3^2+({\hat{\gamma}_3-2\over
2\hat{\gamma}_3})^2\sigma^2=0.
\end{eqnarray}
For $\hat{\gamma}_3=4/3$, the above equation can be greatly
simplified
\begin{equation}
8(1+\sigma)u_3^4-(8\sigma^2+10\sigma+1)u_3^2+\sigma^2=0,
\end{equation}
whose solution is equation (4.11) of KC84.

The total pressure in region 3 can be calculated by $P_{\rm
3,tot}=P_{3}+{u_{4}^2{B'}_{\rm 4}^2 \over 8\pi u_{3}^2}$. The
equality of pressure and velocities along the contact
discontinuity yields $P_{\rm 3,tot}=P_2={4\Gamma_2^2n_1m_{\rm
p}c^2/3},~\Gamma_2=\Gamma_3$. For the Lorentz factor of the
reverse shock (measured by the observer) $\Gamma_{\rm rsh}\gg1$,
$\gamma_{\rm 4}$ can be expressed as $\gamma_{4}\approx
(\eta/\Gamma_{\rm rsh}+\Gamma_{\rm rsh}/\eta)/2$, which in turn
yields $\Gamma_{\rm rsh}\approx (\gamma_4-u_4)\eta$. On the other
hand, $\gamma_{\rm 3}$ can be expressed as $\gamma_{3}\approx
(\Gamma_{2}/\Gamma_{\rm rsh}+\Gamma_{\rm rsh}/\Gamma_{2})/2$,
which in turn yields $\Gamma_{\rm 2}\approx
(\gamma_3+u_3)\Gamma_{\rm rsh}$, where the relation
$\Gamma_2=\Gamma_3$ has been taken. Combing these relations we
have $\Gamma_2\approx (\gamma_3+u_3)(\gamma_4-u_4)\eta$. Finally,
we have the equation (the equality of pressure)
\begin{eqnarray}
{{B'}_{\rm 4}^2\over 4\pi \sigma}\{{u_4 \over
u_3}{(\hat{\gamma}_3-1)\over \hat{\gamma}_3} [{\gamma_4\over
\gamma_3}(1+\sigma(1-Y))-{m_{\rm p}c^2\over \mu_4}]+{\sigma
u_{4}^2 \over 2
u_{3}^2}\}\nonumber\\
={4[(\gamma_3+u_3)(\gamma_4-u_4)\eta]^2n_1m_{\rm p}c^2\over 3},
\end{eqnarray}
where $B'_4=2.6\times 10^{20}{\rm G}~R^{-1}\eta^{-1}$. Equations
(1), (5) and (9) are our basic formulae, with which we can
calculate the dynamical evolution of the magnetized outflow
interacting with the medium, and then the reverse shock emission.
However, these equations cannot be solved analytically unless
$\sigma\gg 1$. For $\sigma\sim 1$ or smaller, only the numerical
calculation can be performed, which is to be presented at the end
of this section.

$R_{\rm cro}$ can be determined as follows: $\beta_{\rm rsh}$, the
velocity of the reverse shock in the observer's frame,  can be
parameterized as (Sari $\&$ Piran 1995)
\begin {equation}
\beta_{\rm rsh}={\Gamma_3n_3\beta_{\rm \Gamma_3}-\Gamma_4
n_4\beta_{\rm \Gamma_4} \over {\Gamma_3 n_3-\Gamma_4 n_4}}.
\end {equation}

Differentially, $(\beta_{\rm \Gamma_4}-\beta_{\rm
rsh})dR=d\Delta$, where $\Delta$ is the width of the reverse shock
penetrating into the ejecta measured by the observer.  $R_{\rm
cro}$ is determined by $\int d\Delta =\int_0^{\rm R_{\rm cro}}
(\beta_{\rm \Gamma_4}-\beta_{\rm rsh})dR=\Delta_0$, where
$\Delta_0$ is the width of the ejecta measured by the observer,
which can be estimated as $\Delta_0\approx cT_{90}/(1+z)$
($T_{90}$ is the observed duration of GRBs).

\subsubsection{The dynamical evolution for $R>R_{\rm cro}$}

After the reverse shock has crossed the ejecta, only the forward
shock exists, whose dynamics have been discussed in great detail
(e.g. Huang et al.  1999; Huang et al. 2000; Panaitescu \& Kumar
2001; Feng et al. 2002). However, in the current work, the ejecta
is magnetized and how to convent the magnetic energy into kinetic
energy is poorly known. But the energy conservation must be
satisfied. As a zeroth order approximation, here we take Huang et
al.'s (1999) differential equation to depict the dynamical
evolution of the magnetized ejecta
\begin{equation}
{d\Gamma \over dm}=-{\Gamma^2-1\over M'_{\rm ej}+\epsilon
m+2(1-\epsilon)\Gamma m}.
\end{equation}
where $\Gamma$ is the bulk Lorentz factor of the outflow,
$\beta=(1-1/\Gamma^2)^{1/2}$ is the corresponding velocity in unit
of light speed; $m$ is the mass of the medium swept by the ejecta,
$dm=4\pi R^2 n_1 m_{\rm p}dR $; $0<\epsilon \leq 1$ is the
radiation efficiency; $M'_{\rm ej}$ is determined by (the energy
conservation) $\gamma_{\rm cro}M'_{\rm ej}c^2=(1-\epsilon_{\rm
rad})E_0$, where $\epsilon_{\rm rad}$ is the fraction of total
energy radiated for $R\leq R_{\rm cro}$, $E_0$ is the initial
isotropic energy of the ejecta. $\gamma_{\rm cro}$ is the bulk
Lorentz factor of the ejecta at $R_{\rm cro}$, which can be
determined by equations (1), (5) and (9).

\subsection{The synchrotron radiation}

\subsubsection{Electron distribution}
In the absence of radiation loss, the distribution of the shock
accelerated electrons behind the blastwave is usually assumed to
be a power law function of electron energy, i.e.,

\begin{equation}
\frac{dN_{\rm e}'}{d\gamma_{\rm e}} \propto \gamma_{\rm e}^{-p},
\,\,\,\,\,\,(\gamma_{\rm e,m}\leq \gamma_{\rm e} \leq\gamma_{\rm
e,M}),
\end{equation}
where $p\approx 2.2$, $\gamma_{\rm e,M}=10^8(B^{\prime}/1{\rm
G})^{-1/2}$ is the maximum Lorentz factor (Dai, Huang \& Lu 1999),
where $B'$ is the comoving frame magnetic field strength. As
usual, for material heated by the forward shock, we assume the
magnetic energy density in the comoving frame is a fraction
$\epsilon_{\rm B}$ of the total thermal energy, $B_2'$ can be
estimated as
\begin{equation}
\frac{B_{\rm 2}'^2}{8\pi}=\epsilon_{\rm
B}(4\gamma'_2+3)(\gamma'_2-1)n_1m_{\rm p}c^2,
\end{equation}
where $\gamma'_2=\Gamma_2$ for $R<R_{\rm cro}$ and
$\gamma'_2=\Gamma$ for $R>R_{\rm cro}$. For region 3 ($R\leq
R_{\rm cro}$), $B'_3=u_4B'_4/u_3$.

Similarly, for region 2 and 3, the thermal energy of electrons in
the comoving frame is assumed to be a fraction $\epsilon_{\rm e}$
of the total thermal energy, then $\gamma_{\rm e,m}$ can be
estimated by\footnote{Here the possible but poorly constrained
$e^\pm$ pair generation during the $\gamma-$ray emission phase and
its impact on the reverse shock emission (e.g. Pilla \& Loeb 1998;
Li et al. 2003; Fan \& Wei 2004; Fan et al. 2004a) have not been
taken into account.}
\begin{equation}
\gamma_{\rm e,m}=\epsilon_{\rm e}\frac{p-2}{p-1}\frac{m_{\rm
p}}{m_{\rm e}}(\gamma'-1),
\end{equation}
where $m_{\rm e}$ is the rest mass of electron,
$\gamma'=\gamma'_2$ for region 2 and $\gamma'=\gamma_{\rm p,th}+1$
for region 3, where the thermal Lorentz factor of the shocked
proton $\gamma_{\rm p,th}$ can be estimated by (see equation (6))
\begin{equation}
\gamma_{\rm p,th}\equiv e_3/n_3m_{\rm p}c^2\approx {1\over
\hat{\gamma}_3}[{\gamma_4\over \gamma_3}(1+\sigma(1-Y))-{m_{\rm
p}c^2\over \mu_4}]{\mu_4 \over m_{\rm p}c^2}.
\end{equation}

It is well known that radiation loss may play an important role in
the process. Sari, Piran \& Narayan (1998) have derived an
equation for the critical electron Lorentz factor, $\gamma_{\rm
c,k}$ ($k=2,3$ represent region 2 and 3), above which synchrotron
radiation is significant
\begin{equation}
\gamma_{\rm c,k}=\frac{12 \pi m_{\rm e} \Gamma'(1-\beta'\mu)
c}{\sigma_{\rm T} B_{\rm k}'^2 t},
\end{equation}
where $t$ is the observer time, $\mu\equiv \cos \Theta$, $\Theta$
is the angle between the velocity of emitting material and the
line of sight; Throughout the rest of this work,
$\Gamma'=\Gamma_2$ ($\beta'$ is the corresponding velocity in unit
of light) in the presence of reverse shock and $\Gamma'=\Gamma$ at
later time. If the emitting material moves on the line of sight,
equation (16) reduces to the familiar form $\gamma_{\rm
c,k}=\frac{6 \pi m_{\rm e} c}{\sigma_{\rm T}\Gamma' B_{\rm k}'^2
t}$. For electrons with Lorentz factors below $\gamma_{\rm c,k}$,
the synchrotron radiation is ineffective.  For electrons above
$\gamma_{\rm c,k}$, they are highly radiative.

In the presence of steady injection of electrons accelerated by
the shock, the distribution of electrons with $\gamma_{\rm
e}>\gamma_{\rm c,k}$ has a power law function with an index of
$p+1$ (Rybicki \& Lightman 1979), while the distribution of
adiabatic electrons is unchanged. The actual distribution should
be given according to the following cases:

\begin{description}
\item (i) For $\gamma_{\rm c}\leq \gamma_{\rm e,m}\leq \gamma_{\rm e,M}$, i.e., the
fast cooling phase
\begin{equation}
 \frac{dN_{\rm e}'}{d\gamma_{\rm e}} =C_1 \left \{
   \begin{array}{ll}
 \gamma_{\rm e}^{-2}\,, \,\,\,\, & (\gamma_{\rm c} \leq \gamma_{\rm e}
                          \leq \gamma_{\rm e,m}), \\
 \gamma_{\rm e,m}^{p-1}\gamma_{\rm e}^{-(p+1)}\,, \,\,\,\, & (\gamma_{\rm e,m}<\gamma_{\rm e}
                          \leq \gamma_{\rm e,M}),
   \end{array}
   \right.
\end{equation}

\begin{equation}
C_1=[{1\over \gamma_{\rm c}}-{p-1\over p}{1\over \gamma_{\rm
e,m}}-{\gamma_{\rm m}^{\rm p-1}\gamma_{\rm e,M }^{\rm -p}\over
p}]^{-1}N_{\rm tot}\,,
\end{equation}
where $N_{\rm tot}$ is the total number of radiating electrons
involved.

\item (ii) For $\gamma_{\rm e,m} < \gamma_{\rm c} \leq \gamma_{\rm
e,M}$, i.e., the slow cooling phase
\begin{equation}
  \frac{dN_{\rm e}'}{d\gamma_{\rm e}} = C_2\left \{
   \begin{array}{ll}
\gamma_{\rm e}^{-p}\,, \,\,\,\, & (\gamma_{\rm e,m} \leq
\gamma_{\rm e}\leq \gamma_{\rm c}), \\
\gamma_{\rm c}\gamma_{\rm e}^{-(p+1)}\,, \,\,\,\, & (\gamma_{\rm
c}<\gamma_{\rm e}\leq \gamma_{\rm e,M}),
   \end{array}
   \right.
\end{equation}
where
\begin{equation}
C_2=[{\gamma_{\rm e,m}^{\rm 1-p}\over p-1}-{\gamma_{\rm c}^{\rm
1-p}\over p(1-p)}-{\gamma_{\rm c}\gamma_{\rm e,M}^{\rm -p}\over
p}]^{-1}N_{\rm tot}.
\end{equation}
\end{description}

As the reverse shock has crossed the ejecta, there are no freshly
heated electrons injected and the ejecta cools adiabatically. Here
we investigate the decay of the magnetic field and the cooling
behavior of electrons. In the current case, the magnetic field is
ordered and satisfies the magnetic flux conservation. As usual,
the ejecta are in the spreading phase and the width can be
estimated as $\Delta'\sim R/\Gamma$. Therefore, $B_3'\propto
R^{-1}\Gamma^{-1}/\Delta'\propto R^{-2}$. The cooling behavior of
electrons is more difficult to estimate. Here we simply assume the
cooling of electrons is not much different from that of the
non-magnetized case. According to M\'{e}sz\'{a}ros \& Rees (1999),
$\gamma_{\rm e,m}\propto R^{\rm -(9-s)/6}$ (so does $\gamma_{\rm c
}$, since all of them cool by adiabatic expansion only), where
$s=0$ for the ISM and $s=2$ for the stellar wind. In the case of
fast cooling, once $\nu_{\rm c,obs}=eB'_3\gamma_{\rm
c}^2\Gamma_3/2\pi m_{\rm e}c$ drops below the observed frequency,
the flux drops exponentially with time (Sari \& Piran 1999). If
the equal arriving time surface has been taken into account, the
flux drops more slowly.

\subsubsection{Relativistic transformations}

In the co-moving frame, synchrotron radiation power at frequency
$\nu '$ from electrons is given by (Rybicki \& Lightman 1979)
\begin{equation}
P'(\nu ') = \frac{\sqrt{3} e^3 B'}{m_{\rm e} c^2}
        \int_{\gamma'_{\rm e}}^{\gamma_{\rm e,M}}
        \left( \frac{dN_{\rm e}'}{d\gamma_{\rm e}} \right)
        F\left(\frac{\nu '}{\nu_{\rm c}'} \right) d\gamma_{\rm e},
\end{equation}
where $e$ is electron charge, $\nu_{\rm c}' = 3 \gamma_{\rm e}^2 e
B' / (4 \pi m_{\rm e} c)$, $\gamma'_{\rm e}=\min\{ \gamma_{\rm e,m
}, \gamma_{\rm c}\}$ and
\begin{equation}
F(x) = x \int_{x}^{+ \infty} K_{5/3}(k) dk,
\end{equation}
with $K_{5/3}(k)$ being the Bessel function. We assume that this
power is radiated isotropically in the comoving frame, $\frac{d
P'(\nu ')}{d \Omega '} = \frac{P'(\nu ')}{4 \pi}$.

The angular distribution of power in the observer's frame is
(Rybicki \& Lightman 1979; see also Huang et al. 2000)
\begin{equation}
\frac{d P(\nu)}{d \Omega} = \frac{1}{\Gamma'^3 (1 - \beta' \mu)^3}
                \frac{dP'(\nu ')}{d \Omega '}
              = \frac{1}{\Gamma'^3 (1 - \beta' \mu)^3}
                \frac{P'(\nu ')}{4 \pi},
\end{equation}
\begin{equation}
\nu = \frac{\nu'}{(1+z)\Gamma' (1 - \beta' \mu)},
\end{equation}
where $z$ is the redshift of the ejecta. Then the observed flux
density at frequency $\nu$ is
\begin{eqnarray}
S_{\nu} &=& \frac{1}{A} \left( \frac{dP(\nu)}{d \Omega}
\frac{A}{D_{\rm L}^2} \right)\nonumber\\
    &=& \frac{1+z}{{\Gamma'}^3 (1 - \beta' \mu)^3} \frac{1}{4 \pi D_{\rm L}^2}
          P'\left((1+z)\Gamma'(1 - \beta' \mu) \nu \right),
\end{eqnarray}
where $A$ is the area of our detector and $D_{\rm L}$ is the
luminosity distance (we assume $H_0=65 \rm km$  $\rm s^{-1}$ $\rm
Mpc^{-1}$, $\Omega_M=0.3$, $\Omega_\wedge=0.7$).

\subsubsection{Equal arrival time surfaces}

Photons received by the detector at a particular time $t$ are not
emitted simultaneously in the burster frame. In order to calculate
observed flux densities, we should integrate over the equal
arrival time surface determined by (e.g. Huang et al. 2000)
\begin{equation}
t = \int \frac{1 - \beta' \mu}{\beta' c} dR \equiv {\rm const},
\end{equation}
within the boundaries.

\subsection{Numerical results}
In our calculation, the number density of the medium (in unit of
$\rm cm^{-3}$) has been taken as
\begin{equation}
 n_1=\left \{
   \begin{array}{ll}
{\rm const.}\,, \,\,\,\, & ({\rm ISM}), \\
 3.0\times 10^{35}A_*R^{-2}\,, \,\,\,\, & ({\rm WIND}),
   \end{array}
   \right.
\end{equation}
respectively, where $A_*=\frac{\dot{M}}{10^{-5}M_\odot~\rm
yr^{-1}}(\frac{v_{\rm w}}{10^{3}{\rm km~s^{-1}}})^{-1}$, $\dot{M}$
is the mass loss rate of the progenitor,  $v_{\rm w}$ is the wind
velocity (Dai \& Lu 1998; Chevalier \& Li 2000). In our numerical
calculation, we take $A_*=1$.

For illustration, we take $E_{\rm kin}=10^{53}{\rm ergs}$, $z=1$,
$L=2\times 10^{51}{\rm ergs~s^{-1}}$ (equally, $T_{90}=50(1+z){\rm
s}$), $\eta=300$, $p=2.2$, $P_4=0$ and $\epsilon_{\rm e}=0.3$. In
the case of ISM-ejecta interaction, the protons in the region 3
are only mild-relativistic or even sub-relativistic, but electrons
are ultra-relativistic, so $\hat{\gamma}_3=13/9$. In the case of
WIND-ejecta interaction, protons heated by the reverse shock are
relativistic, so $\hat{\gamma}_3=4/3$.

\subsubsection{ISM-ejecta interaction case}

The sample very early R-band ($\nu_{\rm R}=4.6\times 10^{14}{\rm
Hz}$) light curves have been shown in figure 1. Before the reverse
shock crosses the ejecta, electrons accelerated by the reverse
shock are in the slow cooling phase and the synchrotron emission
at R-band increases rapidly with time. At tens of seconds after
the main burst, the reverse shock emission at R band is bright to
$m_{\rm R}\sim 10-12{\rm th}$ magnitude and the forward shock
emission is relatively dimmer. Therefore, the very early reverse
shock emission can be detected independently. After the reverse
shock has crossed the ejecta, there are no freshly accelerated
electrons injected and the R-band emission drops sharply.

\begin{figure}
\epsfig{file=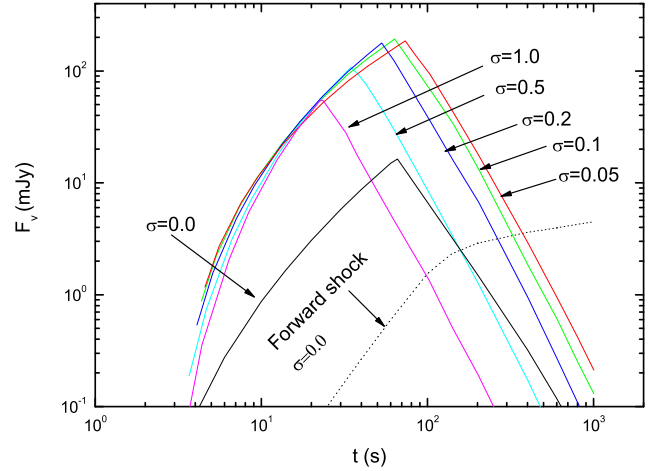, width=100mm} \caption{The very early R-band
($\nu_{\rm R}=4.6\times 10^{14}{\rm Hz}$) light curve powered by
the mildly magnetized outflow (the degrees of the magnetization
have been marked in the figure) interacting with the interstellar
medium. The parameters taken in the calculation are: $z=1$,
$E_{\rm kin}=10^{53}{\rm ergs}$, $L=2\times 10^{51}{\rm
ergs~s^{-1}}$, $\eta=300$, $n_1=1{\rm cm^{-3}}$, $\epsilon_{\rm
e}=0.3$ and the radiation efficiency $\epsilon=\epsilon_{\rm e}$.
For $\sigma=0$ and the forward shock, it is assumed that
$\epsilon_{\rm B}=0.01$.}
\end{figure}

In figure 1, there are two interesting phenomena: (i) The peak
flux at R-band increases with the increasing $\sigma$ for
$\sigma<0.1$, but for $\sigma>0.1$ the peak flux at R-band
decreases with the increasing $\sigma$ (A similar result has been
obtained by Zhang \& Kobayashi 2004). This behavior can be
understood as follows: For $\sigma\ll 1$, the electrons heated by
the reverse shock is in slow cooling phase. At $R_{\rm cro}$, the
typical synchrotron radiation frequency $\nu_{\rm m,obs}=\Gamma_2
\gamma_{\rm e,m}^2eB'_3/[2\pi (1+z) m_{\rm e}c]$ is much lower
than $\nu_{\rm R}$, as is $\nu_{\rm c,obs}$. We have approximately
$F_{{\rm \nu_{\rm R}}}\propto {B'_3}^{(p-1)/2}$. Roughly speaking,
the observed flux increases with the increasing $\sigma$.  For
larger $\sigma$, the reverse shock has been suppressed and the
electrons involved in the emission decrease. So $F_{\rm \nu_{\rm
R}}$ drops again (see Zhang \& Kobayashi 2004 for more detailed
explanation). (ii) For $\sigma\sim 1$, the crossing time $t_{\rm
cro}$ (at which the reverse shock crosses the ejecta; in figures 1
and 2, it equals the peak time of the reverse shock emission) is
much shorter than $T_{90}$. Here we explain this in some detail.
In the presence of reverse shock, differentially, $t_{\rm cro}$
satisfies \footnote{If the reverse shock is Newtonian, i.e.,
$\Gamma_{\rm rsh}\approx (1-f)\Gamma_4$, where $0< f \ll 1$. We
have $\beta_{\Gamma_4}-\beta_{\rm rsh}\approx {f\over
\Gamma_4^2}$. Equation (28) should be written into $dt_{\rm
cro}\approx {1+z\over 2f}{d\Delta\over c}$. Approximately, $t_{\rm
cro}\approx {T_{90}\over 2f}\gg T_{90}.$}
\begin{equation}
dt_{\rm cro}=(1+z)\frac{1-\beta_{\rm \Gamma_3}} {\beta_{\rm
\Gamma_4}-\beta_{\rm rsh}} \frac{d\Delta}{c} \approx (1+z)
\frac{\Gamma_{\rm rsh}^2}{\Gamma_3^2}\frac{d\Delta}{c}.
\end{equation}
With $\Gamma_{\rm rsh}\approx (\gamma_3-u_3)\Gamma_3$, we have
$dt_{\rm cro} \approx
\frac{1+z}{(\gamma_3+u_3)^2}\frac{d\Delta}{c}$. Approximately
\begin{equation}
t_{\rm cro}=\int{dt_{\rm cro}}\approx \frac{1}{(\gamma_{\rm
3,cro}+u_{\rm 3,cro})^2}T_{90},
\end{equation}
where $\gamma_{\rm 3,cro}$ and $u_{\rm 3,cro}$ are the corresponding value of
$\gamma_3$ and $u_3$ at $R_{\rm cro}$. In the case of non-magnetization and the
reverse shock is relativistic, $\gamma_{\rm 3,cro}\approx \sqrt{9/8}$, which
yields $t_{\rm cros}\approx (1+z)\Delta_0/2c\approx T_{\rm 90}/2$ (This result
coincides with the ``classical'' result of Sari \& Piran 1995). While for
$\sigma=1$, $\gamma_{\rm 3,cro}\approx 1.37$ which results in $t_{\rm
cros}\approx T_{\rm 90}/5$, which is quite consistent with our numerical result
in Fig.1. Equation (29) applies to the following WIND-ejecta interaction case as
well, if only the reverse shock is relativistic or at least mild-relativistic.

\subsubsection{WIND-ejecta interaction case}

In the case of WIND-ejecta interaction (see figure 2), before the reverse shock
crosses the ejecta, the electrons accelerated by the reverse shock are in fast
cooling phase and the R-band emission increases only slightly with time. This
temporal behavior is very similar to that of non-magnetized fireball case (See
Wu et al. (2003) for an analytical investigation). At $t_{\rm cro}$, the reverse
shock emission at R band is very bright ($m_{\rm R}\sim 9-10{\rm th}$ magnitude)
and the forward shock emission is relatively dimmer. Therefore, the very early
reverse shock emission can be detected independently, too. After the reverse
shock has crossed the ejecta, the R-band emission drops sharply.

\begin{figure}
\epsfig{file=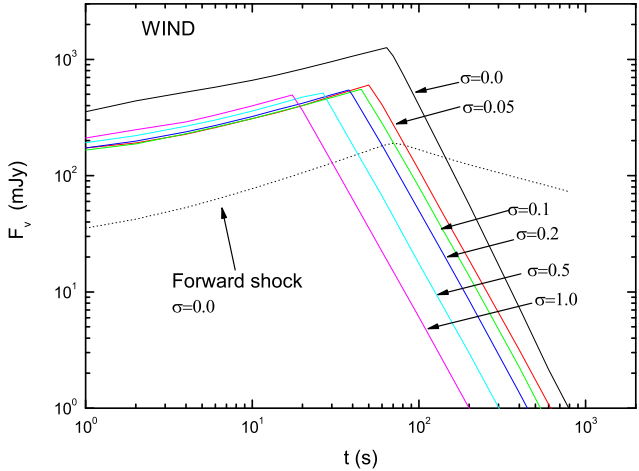, width=100mm} \caption{The very early R-band
light curve powered by the mildly magnetized outflow (the degree
of the magnetization has been marked in the figure) interacting
with the stellar wind. The parameters taken here are the same to
those of figure 1 except $n_1=3\times 10^{35}R^{-2}{\rm
cm^{-3}}$.}
\end{figure}

Since the current reverse shock is relativistic, as implied by
equation (29), $t_{\rm cro}$ decreases with increasing $\sigma$.
However, in figure 2, the R-band reverse shock emission is
brightest at $\sigma=0$, which seems to be inconsonant with the
result shown in figure 1. The main reason for this ``divergence''
is: For $R<10^{17}{\rm cm}$, the WIND is far denser than the ISM.
Consequently, the reverse shock is very strong and the ejecta has
been decelerated significantly at a radius $\sim 3\times
10^{15}{\rm cm}$ (Note that in the case of ISM-ejecta interaction,
the corresponding radius is $\sim 10^{17}{\rm cm}$). Even for
$\sigma=0$, the electrons heated by the reverse shock are in the
fast cooling phase and $\nu_{\rm m,obs}$ is much higher than the
observer frequency $\nu_{\rm R}$. With the increasing $\sigma$,
the corresponding $R_{\rm cro}$ decreases. As a result, $B'_3$
increases. However, now $F_{\rm \nu_{\rm R}}\propto
{B'_3}^{-1/2}$. Consequently, the reverse shock emission powered
by the high $\sigma$ ejecta interacting with WIND is dimmer than
that powered by the low $\sigma$ ones.

\section{The linear polarization}
It is well known that for the totally ordered magnetic
configuration, high linear polarization is expected. For the
random magnetic configuration, mild linear polarization can be
expected if some specific geometry effects have been taken into
account. Sometimes the magnetic field is not only ordered or only
random. Then it is interesting to investigate its linear
polarization. Here we propose a simple formula to describe the
linear polarization properties of a slab of such a mixed magnetic
field, with which we can see the impact of the ordered field on
the polarization.

Following Laing (1980; See his Appendix A1 for detail), the
coordinates involved are defined as follows ( see figure 3):

\begin{figure}
\epsfig{file=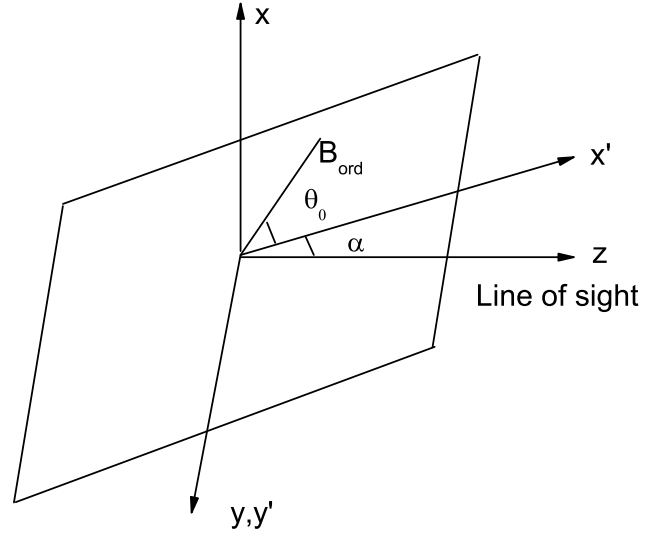, width=100mm} \caption{Coordinates used in the
calculation of the polarization properties of a slab of mixed
field (after Laing 1980).}
\end{figure}

$\alpha$ is the angle between the plane of the ejecta and the line
of sight; $x$, $y$, $z$ are rectangular coordinates with the
z-axis pointing towards the observer (i.e., the direction
$\textbf{n}$) and the y-axis parallel to the ``local'' plane of
the ejecta; $x'$, $y'$ are coordinates in the plane of the ejecta,
$y'$ is parallel to y; $\theta$ is the angle between the field
direction and the $x'$ axis at any point in the ejecta; $\chi$ is
the position angle of the E-vector of the polarized radiation,
measured from the $x-y$ plane. Therefore the random (ordered)
magnetic-field vector $\textbf{B}_{\rm ran}$ ($\textbf{B}_{\rm
ord}$) at a point in the slab are\footnote{According to Medvedev
\& Loeb (1999), the configuration of the magnetic field generated
in shocks is tangled with the front of shock surface. As a result,
locally, in a finite scale, it is planar.} $\textbf{B}_{\rm
ran}=B(\cos \theta \sin \alpha,~\sin \theta,~\cos\theta \cos
\alpha )$, $\textbf{B}_{\rm ord}=B_0(\cos \theta_0 \sin
\alpha,~\sin \theta_0,~\cos\theta_0 \cos \alpha )$ respectively.
Thus the total magnetic field is
\begin{eqnarray}
\textbf{B}=B[(\cos \theta+b\cos \theta_0) \sin \alpha,~\sin
\theta +b\sin\theta_0,~\nonumber\\
(\cos \theta +b\cos\theta_0) \cos \alpha],
\end{eqnarray}
where $b\equiv B_0/B$. The electric field of a linearly polarized
electromagnetic wave is directed along the vector
\begin{eqnarray}
\textbf{e}&=&\textbf{n}\times \textbf{B}\nonumber\\
&=&[-(\sin \theta+b\sin \theta_0),~(\cos \theta +b\cos
\theta_0)\sin \alpha,~0].
\end{eqnarray}
Now $\chi$ satisfies
\begin{equation}
\tan\chi=-\frac{\sin \alpha(\cos \theta+b\cos \theta_0)}{\sin
\theta+b\sin \theta_0}.
\end{equation}
With equation (32), it is easy to get expressions for $\cos
(2\chi)$ and $\sin(2\chi)$. With equations (1)--(3) of Laing
(1980) and assuming the spectra index $p=3$, we have the Stokes
parameters
\begin{equation}
Q={3\over 4}[\cos^2{\alpha}+2b^2(\sin^2
\theta_0-\sin^2\alpha\cos^2\theta_0)]\pi B^2,
\end{equation}
\begin{equation}
U=1.5\sin \alpha \sin 2\theta_0 \pi b^2 B^2,
\end{equation}
\begin{equation}
I={(1+\sin^2\alpha)+2b^2( \sin^2\theta_0+\sin^2\alpha
\cos^2\theta_0)}\pi B^2.
\end{equation}
For $\theta_0=\pi/2 ~{\rm or}~3\pi/2$, we have $U=0$ and the
degree of the linear polarization
\begin{equation}
\Pi \equiv{Q\over I}={3\over
4}\frac{\cos^2\alpha+2b^2}{(1+\sin^2\alpha)+2b^2}\geq {3\over
4}\frac{b^2}{1+b^2}.
\end{equation}

For $b\gg1$, $\Pi$ has the maximum value $\Pi={3\over 4}$. In the
current work, $\sigma\sim 0.05-1$, the corresponding toroidal
magnetic field is far stronger than that generated in reverse
shock, i.e., $b\gg1$, so the local point polarization can be as
high as $75\%$. For ultra-relativistic ejecta, due to the beaming
effect, only the emission coming from a very tight cone around the
line of sight can be detected. If the line of sight is slightly
off the symmetric axis of the ordered magnetic field, the
orientation of the viewed magnetic field is nearly the same. The
high linear net polarization is expected since the local high
linear polarization cannot be averaged effectively. The detailed
numerical calculation of the net polarization will be presented
elsewhere.

Here, for simplicity, following the treatment of Granot \& K\"{o}nigl (2003),
the net Stokes parameters of the ordered magnetic field ($U_{\rm ord},~Q_{\rm
ord},~I_{\rm ord}$) and of the random magnetic field ($U_{\rm ran},~Q_{\rm
ran},~I_{\rm ran}$) are calculated separately. Therefore

\begin{eqnarray}
\Pi_{\rm net}=\frac{Q_{\rm ran}+Q_{\rm ord}}{I_{\rm ran}+I_{\rm
ord}} =\frac{0+AI_{\rm ord}}{I_{\rm ran}+I_{\rm
ord}}=0.60{b^2\over 1+b^2},
\end{eqnarray}
where $Q_{\rm ran}=U_{\rm ran}=U_{\rm ord}=0$ for the symmetrical
viewing area, and $I\propto B^2$ for $p=3$. For $b\rightarrow
\infty$\footnote{In Lyutikov, Pariev \& Blandford (2003), the
Stokes parameters are pulse-integrated and the resulting $\Pi_{\rm
net}=0.56$. Considered that the photons emitted at the same time
but with different angles arrive at different time, a more
detailed calculation suggests $\Pi_{\rm net}=0.64$. Such a
difference is not large. Therefore, and partly for simplicity, we
take $\Pi_{\rm net}=0.60$.}, $\Pi_{\rm net}\simeq 0.60$, so we
take $A=0.60$.

Equation (37) is favored by the fact that for $b=1$, $1/\sqrt{3}$
and 0, it gives $\Pi_{\rm net}=0.30$, 0.15 and 0 respectively,
which coincides with the result of Granot \& K\"{o}nigl (2003)
excellently. Then we believe that equation (37) provides us a
rough but reliable estimation on the impact of the ordered
magnetic field on the linear polarization. Equation (37) is valid
only when the viewed emitting region for the random magnetic field
is axisymmetric. If it is not, the random magnetic field may play
an important role. The detailed calculation for that case is
beyond the scope of this paper.

\section{Discussion and conclusion}

The reverse shock emission in the framework of the standard
fireball model of GRBs has been discussed in great detail (e.g.
Sari \& Piran 1999; M\'{e}sz\'{a}ros \& Rees 1999; Wang et al.
2000; Kobayashi 2000; Wu et al. 2003; Zhang, Kobayashi \&
M\'{e}sz\'{a}ros 2003; Nakar \& Piran 2004). The very early
afterglow of the X-ray Flashes has been investigated by Fan et al.
(2004b) recently. For typical parameters and reasonable
assumptions about the velocity of the source expansion, a strong
optical flash ${\rm m_{\rm R}}\approx 9-17{\rm th}$ magnitude is
expected (e.g. Sari \& Piran 1999; Wu et al. 2003; Fan et al.
2004b). However, despite intensive efforts, only three candidates
(GRB 990123, GRB 021004 and GRB 021211) have been reported (Sari
\& Piran 1999; Kobayashi \& Zhang 2003; Wei 2003 and references
listed therein). It is unclear why. Interestingly, modeling the
reverse shock emission of GRB 990123 and GRB 021211 suggests that
the reverse shock emission region is magnetized---In other words,
the magnetic energy density in region 3 is far stronger than that
in region 2 (Fan et al. 2002; Zhang, Kobayashi \& M\'{e}sz\'{a}ros
2003). There are two possible explanations: One is that the
magnetic field coming from the central source has been dissipated
significantly, i.e., the case considered in this paper. The other
is that the magnetic field is generated in internal shock. In the
internal shock model, the generated magnetic field can be as high
as $0.01-0.1$ times the total thermal energy of the shocked
baryons. The generated magnetic field is randomly oriented in
space, but always lies in the plane of the shock front, for which
the jump condition derived in $\S3$ is satisfied in the coherence
scale. More importantly, the annihilation timescale for the random
magnetic field is much longer than the dynamical timescale of the
fireball, then the existing of the generated magnetic field can
affect the very early afterglow (Medvedev \& Loeb 1999). However,
the coherence scale of the generated magnetic field is so small
($\sim 10^3~{\rm cm}$) that there is no net polarization in the
early multi-wavelength emission unless some geometry effects have
been taken into account (e.g. Medvedev \& Loeb 1999). However, as
shown in $\S4$, if part of the magnetic field is ordered, high
linear polarization can be detected (see also Granot \& K\"{o}nigl
2003). Therefore polarization detection at very early times may
provide us the chance to distinguish between the usual baryon-rich
fireball model and the Poynting flux-dominated outflow model for
GRBs.

The predicted very early afterglow in the R band is bright to
$m_{\rm R}\sim 10-12{\rm th}$ magnitude, which is strong enough to
be detected by current telescopes, such as the ROTSE-IIIa
telescope system, which is a 0.45-m robotic reflecting telescope
and managed by a fully-automated system of interacting daemons
within a Linux environment. The telescope has an f-ratio of 1.9,
yielding a field of view of 1.8${\rm \times}$1.8 degrees. The
control system is connected via a TCP/IP socket to the Gamma-ray
Burst Coordinate Network (GCN), which can respond to GRB alerts
fast enough ($<10{\rm s}$). ROSTE-IIIa can reach 17th magnitude in
a 5-s exposure, 17.5 in 20-s exposure (see Smith et al. 2003 for
details). Another important instrument for detecting the very
early afterglow is the Ultraviolet and Optical Telescope (UVOT) on
board the Swift Satellite. The Burst Alert Telescope (BAT) is
another important telescope, with which hundreds of bursts per
year to better than 4 arc minutes location accuracy will be
observed. Using this prompt burst location information, Swift can
slew quickly to point the on-board UVOT at the burst for continued
afterglow studies. The spacecraft's 20$-$70 second time-to-target
means that about $\sim$100 GRBs per year (about 1/3 of the total)
will be observed by the narrow field instruments during
$\gamma-$ray emission phase. The UVOT is sensitive to magnitude 24
in a 1000 second exposure (For a linear increase of the
sensitivity with the exposure time, that means a sensitivity of
magnitude 19 in a 10 second exposure). These two telescopes are
sufficient to detect the very early optical emission predicted
here.

In this work, the problem has been treated under the ideal MHD
limit. In fact, magnetic dissipation may play a role (e.g. Fan et
al. 2004c). Thus our treatment is a simplification of the real
situation, and further considerations are needed to fully depict
the physics involved.

%%%%%%%%%%%%%%%%%%%%%%%%%%%%%%%%%
\acknowledgements  We thank T. Lu, Z. G. Dai, Y. F. Huang, X. Y.
Wang \& X. F. Wu  for fruitful discussions. This work is supported
by the National Natural Science Foundation (grants 10073022,
10225314 and 10233010), the National 973 Project on Fundamental
Researches of China (NKBRSF G19990754).

\end{document}